\documentclass{elsart}

\usepackage{natbib}

\usepackage{epsfig}

\usepackage{amssymb}

\begin{document}

\begin{frontmatter}



\title{Gamma-ray burst observations with H.E.S.S.}


\author[LSW]{Pak-Hin Tam}
\author[Durham]{Paula Chadwick}
\author[Montpellier]{Yves Gallant}
\author[Tuebingen]{Dieter Horns}
\author[LSW]{Gerd P\"{u}hlhofer}
\author[Australia]{Gavin Rowell}
\author[LSW]{Stefan Wagner}
\collab{for the H.E.S.S. collaboration}

\address[LSW]{Landessternwarte, Universit\"{a}t Heidelberg, Germany}
\address[Durham]{University of Durham, Department of Physics, U.K.}
\address[Montpellier]{Laboratoire de Physique Th\'eorique et Astroparticules, IN2P3/CNRS, Universit\'e
Montpellier II, France}
\address[Tuebingen]{Institut f\"ur Astronomie und Astrophysik, Universit\"at T\"ubingen, Germany}
\address[Australia]{School of Chemistry \& Physics, University of Adelaide, Australia}

\begin{abstract}
The High Energy Stereoscopic System (H.E.S.S.) consists of four Imaging Atmospheric Cherenkov Telescopes (IACTs) in Namibia for the detection of cosmic very-high-energy (VHE) gamma-rays. Gamma-ray bursts (GRBs) are among the potential VHE gamma-ray sources. VHE $\gamma$-emission from GRBs is predicted by many GRB models. Because of its generally fast-fading nature in many wavebands, the time evolution of any VHE $\gamma$-radiation is still unknown. 
In order to probe the largely unexplored VHE $\gamma$-ray spectra of GRBs, a GRB observing program has been set up by the H.E.S.S. collaboration. With the high sensitivity of the H.E.S.S. array, VHE $\gamma$-ray flux levels predicted by GRB models are well within reach. We report the H.E.S.S. observations of and results from some of the reported GRB positions during March 2003 -- May 2006.

\end{abstract}


\end{frontmatter}

\section{H.E.S.S. telescopes}

The H.E.S.S. array is a system of four 13m-diameter IACTs located in the Khomas Highland of Namibia~\citep{hinton04}. Since the completion of the whole array in late 2003, H.E.S.S. has proven to be very successful in VHE $\gamma$-ray astronomy, thus opening a new era in astronomy in this observational window. For a point source with integral flux $\sim1.4\times10^{-11}$~ph~cm$^{-2}$s$^{-1}$ above 1~TeV and spectral index 2.6, only a 2-hour H.E.S.S. observation is required for a 5$\sigma$ detection. With this high sensitivity, we are capable to detect any signal comparable to that predicted in~\citet{zhang01} up to several days (see next section). A review of the system and observational highlights of H.E.S.S. can be found in~\citet{hofmann05}.

\section{Very-high-energy afterglow emission from GRBs}

The highest energy radiation from GRBs ever detected unambiguously
was a $\sim18$~GeV photon coming from
GRB~940217 using EGRET 1.5 hour after the GRB onset~\citep{hurley94}. There is also no evidence of high-energy cut-off in the spectra of seven GRBs detected with EGRET at energies $>30$~MeV. There
could be an energy flux from GRBs in the largely unexplored VHE $\gamma$-ray
regime comparable to that radiated in keV-MeV or X-ray-to-radio
energies.

In the context of standard models, photons with energies up to $\sim10$~TeV from GRBs are expected. Possible radiation mechanisms for VHE $\gamma$-ray production include electron Inverse-Compton (IC) emission, proton synchrotron radiation, and $\pi_0$--decay from $p\gamma$ interactions. In one case considered by \citet{zhang01} where electron IC emission dominates (Fig. 2b in the reference), an energy flux of about $5\times10^{-11}$~erg~cm$^{-2}$s$^{-1}$ at 1~TeV one day after GRB onset is predicted, if one assumes a redshift of 0.15. This is well within H.E.S.S. detection limit. 
The detection of VHE $\gamma$ photons (and its quantity) or upper limits could be used to constrain GRB properties, eg. bulk Lorentz factor and ambient density~\citep{peer05,wang05}.

At cosmological distances, one has to take into account the absorption of VHE $\gamma$ photons by extragalactic background light (EBL; their density in the range of infrared to optical is still uncertain). However for low-redshift GRBs and sub-TeV energies, the attenuation is less significant. There are also evidences from distant blazar spectra that the Universe is more transparent for VHE $\gamma$ photons than previously thought~\citep{aha06a}. Thus, current air Cherenkov systems are able to observe out to $z\sim1$ at $\sim100$~GeV.

\section{H.E.S.S. GRB observing program}


We currently follow on-board GRB triggers distributed by {\it
Swift}, as well as triggers from HETE~II and INTEGRAL confirmed by
ground-based analysis. Upon the reception of a GRB Coordinates Network (GCN) notice from one of these satellites (with good indications of being a true GRB
), we will observe the burst position as soon as possible, limited to $\mathrm{ZA} < 45$~degrees (for reasonably low energy threshold) and HESS dark time\footnote{H.E.S.S. observations are taken in darkness and when the moon is below horizon; the dark time fraction is therefore about 0.2}. We start observing the burst position up to 24 hours after the burst time.



We have been observing GRBs since early 2003. At the beginning of
2005, a GRB coordination team was set up and since then our GRB
observation program has been fully established. By May
2006, 14 GRB positions had been observed using H.E.S.S. (see
Table~\ref{GRBtable}). The bursts are ranked according to the relative expected VHE $\gamma$-signal as estimated from the fluence in the 15-150 keV band multiplied by a factor of $t^{-1.3}$, where $t$ is the delay observation time. For simplicity, the effect of EBL absorption is neglected here.

\section{Data Analysis and Results}
Calibration of data, the event reconstruction and rejection of the cosmic-ray background (i.e. $\gamma$-ray event selection criteria) were performed as described in~\citet{aha06b}. Except for the case of GRB~030329, where a different analysis cut was used because only two telescopes were operating, standard analysis cuts as described in~\citet{aha06b} were applied to each GRB to search for any possible signal.

No evidence of excess events for any GRB observed using H.E.S.S. was seen. The 99.9\%
confidence level (c.l.) upper limits using the method of~\citet{feldman98}
for each GRB are included in Table~\ref{GRBtable}. No EBL correction was applied to the upper limits shown here.

\begin{table*}
\caption{GRBs observed with H.E.S.S. from March 2003 to May 2006, ranked according to the relative expected VHE $\gamma$-signal (see text). For each burst, start observation time, live time, mean zenith angle (ZA), energy threshold ($E_{\rm th}$) and 99.9\% c.l. upper limits (ULs) of the observation are shown.}
\label{GRBtable}
\begin{center}
\begin{tabular}{lcccccc}
    \hline
      GRB     & Observation starts  & live time & mean ZA & $E_{\rm th}$ & ULs ($>E_{\rm th}$) & redshift (z)     \\
              & after GRB onset     & (hrs)     & (deg)   & (GeV)        & (cm$^{-2}$ s$^{-1}$) & \\
    \hline
      050922C & 52 min              & 0.7      & 23      & 200          & $1.22\times10^{-11}$& 2.199$^\mathrm{a}$      \\
      050801  & 16 min              & 0.5      & 43      & 370          & $3.40\times10^{-12}$& 1.56$^\mathrm{b}$       \\
      041211  & 9.5 h               & 2.0      & 46      & 420          & $4.00\times10^{-12}$& --             \\
      041006  & 10.4 h              & 1.4      & 27      & 220          & $1.01\times10^{-11}$& 0.716$^\mathrm{c}$      \\
      040425  & 26 h                & 0.4      & 28      & 230          & $2.37\times10^{-11}$& --             \\
      030821  & 18 h                & 1.0      & 28      & 290          & $1.52\times10^{-11}$& --             \\
      060526  & 4.7 h               & 1.9      & 25      & 200          & $5.90\times10^{-12}$& 3.21$^\mathrm{d}$       \\
      030329  & 11.5 d              & 0.5      & 60      & 1400         & $2.58\times10^{-12}$& 0.169$^\mathrm{e}$      \\
      050209  & 20.2 h              & 2.5      & 48      & 520          & $3.32\times10^{-12}$& --             \\
      050726  & 10.8 h              & 2.0      & 40      & 400          & $4.22\times10^{-12}$& --             \\
      060403  & 13.6 h              & 0.9      & 39      & 310          & $9.37\times10^{-12}$& --             \\
      050607  & 14.8 h              & 1.5      & 37      & 290          & $5.39\times10^{-12}$& --             \\
      060505  & 19.4 h              & 2.0      & 42      & 450          & $6.29\times10^{-12}$& 0.089$^\mathrm{f}$      \\
      050509C & 21 h                & 1.0      & 22      & 220          & $1.08\times10^{-11}$& --             \\
      \hline
  \end{tabular}
\end{center}

\vspace*{.6cm}
\noindent
$^\mathrm{a}$\citet{delia05}; $^\mathrm{b}$Photometric z according to \citet{pasquale07}; $^\mathrm{c}$\citet{price04}; $^\mathrm{d}$\citet{berger06}; $^\mathrm{e}$\citet{stanek03}; $^\mathrm{f}$\citet{ofek06}
\end{table*}


\section*{Acknowledgements}
P.H. Tam acknowledges financial support from the International Max Planck Research School for Astronomy and Cosmic Physics at the University of Heidelberg. PHT, GP, and SW acknowledge support by BMBF/DESY through grant 05CH5VH/0.


\begin{thebibliography}{}


\bibitem[Aharonian et al.(2006a)]{aha06a} Aharonian,~F. et al. (HESS collaboration) 2006a, {\em Nature}, {\bf 440}, 1018.
\bibitem[Aharonian et al.(2006b)]{aha06b} Aharonian,~F. et al. (HESS collaboration) 2006b, {\em A\&A}, {\bf 457}, 899.
\bibitem[Berger and Gladders(2006)]{berger06} Berger,~E. and Gladders,~M. 2006, {\em GCN Circular}, 5170.
\bibitem[D'Elia et al.(2005)]{delia05} D'Elia,~V. et al. 2005, {\em GCN Circular} 4044.
\bibitem[De Pasquale et al.(2007)]{pasquale07} De Pasquale,~M. et al. 2007, {\em M.N.R.A.S.}, in press
\bibitem[Feldman and Cousins(1998)]{feldman98} Feldman,~G.~J. and Cousins,~R.~D. 1998, {\em Phys. Rev. D.}, {\bf 57}, 3873.
\bibitem[Hinton(2004)]{hinton04}
    Hinton,~J.~A. 2004, {\em New Astronomy Review}, {\bf 48}, 331.
\bibitem[Hofmann(2005)]{hofmann05} Hofmann,~W. 2005, H.E.S.S. status {\em in Proc. Conf. Towards a Network of Atmospheric Cherenkov Detectors VII, Palaiseau, France}, 43.
\bibitem[Hurley et al.(1994)]{hurley94} Hurley,~K. et al. 1994, {\em Nature}, {\bf 372}, 652.
\bibitem[Ofek et al.(2006)]{ofek06} Ofek,~E.~O. et al. 2006, {\em GCN Circular}, 5123.
\bibitem[Pe'er and Waxman(2005)]{peer05} Pe'er, A. and Waxman, E. 2005, {\em Ap. J.}, {\bf 633}, 1018.
\bibitem[Price et al.(2004)]{price04} Price,~P.~A. et al. 2004, {\em GCN Circular}, 2791.
\bibitem[Stanek et al.(2003)]{stanek03} Stanek,~K.~Z. et al. 2003, {\em Ap. J.}, {\bf 591}, L17.
\bibitem[Wang et al.(2005)]{wang05} Wang~X.~Y. et al. 2005, {\em A\&A}, {\bf 439}, 957.
\bibitem[Zhang and M\'esz\'aros(2001)]{zhang01}
    Zhang,~B. and M\'esz\'aros,~P. 2001, {\em Ap. J.}, {\bf 559}, 110.




\end{thebibliography}
\end{document}